\documentclass[aps,noshowpacs,tightenlines,nobalancelastpage,floatfix,twocolumn]{revtex4}

\usepackage{amsmath}
\usepackage{amssymb}
\usepackage{graphicx}
\usepackage{dcolumn}
\usepackage{bm}
\usepackage{latexsym}
\usepackage{graphicx}
\usepackage{color}
\usepackage{draftcopy}

\begin{document}
\title{Observation of coherent many-body Rabi oscillations}

\date{\today }
\author{Y. O. Dudin, L. Li, F. Bariani, and A. Kuzmich}
\affiliation{School of Physics, Georgia Institute of Technology, Atlanta, Georgia
30332-0430}

\maketitle
\textbf{A two-level quantum system coherently driven by a resonant electromagnetic field oscillates sinusoidally between the two levels at frequency $\Omega$ which is proportional to the field amplitude \cite{griffiths}. This phenomenon, known as the Rabi oscillation, has been at the heart of atomic, molecular and optical physics since the seminal work of its namesake and coauthors \cite{rabi}. Notably, Rabi oscillations in isolated single atoms or dilute gases form the basis for metrological applications such as atomic clocks and precision measurements of physical constants \cite{ramsey}. Both inhomogeneous distribution of coupling strength to the field and interactions between individual atoms reduce the visibility of the oscillation and may even suppress it completely. A remarkable transformation takes place in the limit where only a single excitation can be present in the sample due to either initial conditions or atomic interactions: there arises a collective, many-body Rabi oscillation at a frequency $\sqrt{N}\Omega$ involving all $N \gg 1$ atoms in the sample \cite{dicke}. This is true even for inhomogeneous atom-field coupling distributions, where single-atom Rabi oscillations may be invisible. When one of the two levels is a strongly interacting Rydberg level, many-body Rabi oscillations emerge as a consequence of the Rydberg excitation blockade. Lukin and coauthors outlined an approach to quantum information processing based on this effect \cite{lukin}. Here we report initial observations of coherent many-body Rabi oscillations between the ground level and a Rydberg level using several hundred cold rubidium atoms. The strongly pronounced oscillations indicate a nearly complete excitation blockade of the entire mesoscopic ensemble by a single excited atom. The results pave the way towards quantum computation and simulation using ensembles of atoms.}

\begin{figure*}[thb]
\includegraphics[scale=1]{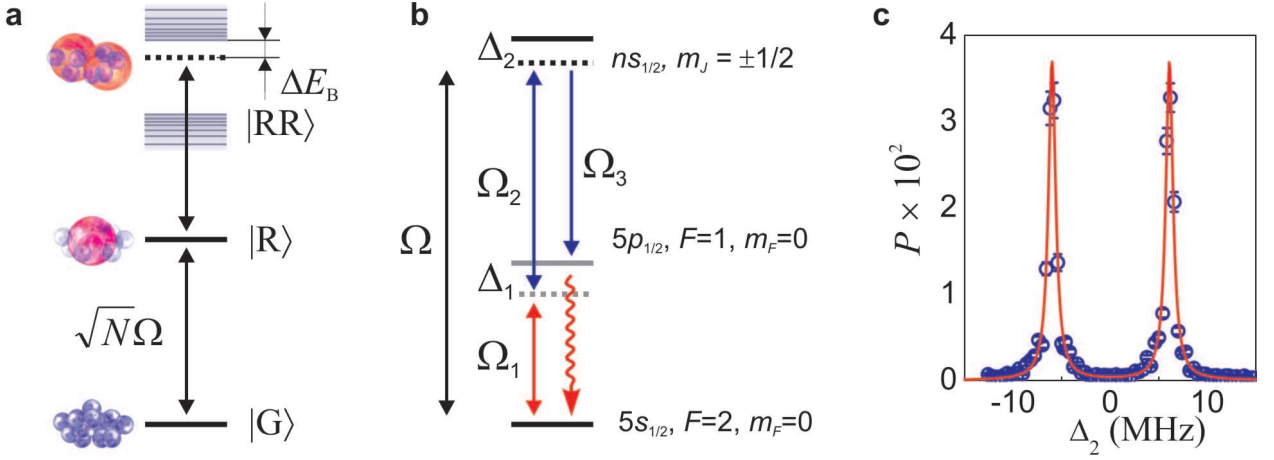}
\caption{\sffamily{  \textbf{Rydberg excitation of an atomic ensemble. a,} Illustration of the excitation blockade of more than one Rydberg atom in the ensemble. The coherent laser driving couples the collective ground state $|G\rangle$ and the state with one Rydberg atom $|R\rangle$ with Rabi frequency $\sqrt{N}\Omega$. The doubly excited states $|RR\rangle$ are shifted in energy out of laser resonance by the strong atomic interactions. {\bf b,} Single-atom energy levels for $^{87}$Rb. Electronic, hyperfine, and Zeeman quantum numbers are shown. The detuning from the intermediate $|5p_{1/2}\rangle$ level is $\Delta_{1}=-40$ MHz. {\bf c,} Probability $P$ of photoelectric detection event per trial as a function of two-photon detuning $\Delta_2$ for level $|102s_{1/2}\rangle$. It shows the two $m_j=\pm1/2$ Zeeman components split by the bias magnetic field. The solid curve is a sum of two Lorentzian functions fit with the 0.9 MHz widths (fwhm) of the peaks determined by the 1 $\mu$s excitation duration.
} } \label{fig:setup}
\end{figure*}

A two-level quantum system coherently driven by a quasi-resonant electromagnetic field is one of the centerpieces of modern quantum physics \cite{griffiths,ramsey}. A wide array of two-level systems have been realized, with atoms, molecules, nuclei, and Josephson junctions being some of the prominent settings. More than half a century ago Dicke recognized that an atomic ensemble coupled to an electromagnetic field cannot always be treated as a collection of independent atoms \cite{dicke}. His ground-breaking work gave rise to a rich field of collective atom-field interaction physics \cite{allen}.

A key prediction of Dicke's theory is that under certain conditions atom-field coupling is enhanced by a factor $\sim \sqrt{N}$ when compared to one atom. Collectively-enhanced atom-field coupling has since been observed in a variety of settings involving either the emission or absorption of radiation.
A coherent multi-atom Rabi oscillation at a frequency $\sqrt{N}\Omega$ is a particularly dramatic manifestation of quantum mechanics at work on mesoscopic scales, where an entire ensemble exhibits the dynamical behavior of a single two-level system. In 2001, Lukin {\it et al.} proposed to realize many-body Rabi oscillations in ensembles of atoms driven by a laser tuned to a Rydberg level, and outlined designs for scalable quantum gates for quantum computation and simulation and generation of entangled collective states for metrology beyond the standard quantum limit.

 \begin{figure*}[bth]
\includegraphics[scale=0.8]{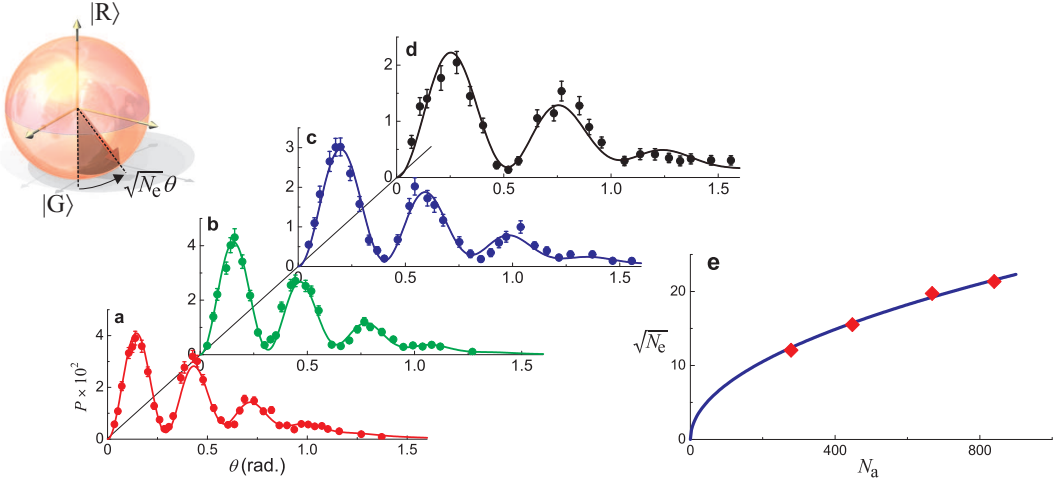}
\caption{\sffamily{ \textbf{Coherent many-body Rabi oscillations of a mesoscopic atomic ensemble.} In panels (a)-(d), probability of photoelectric detection $P$ as a function of the single-atom Rabi angle $\theta$ is shown; upper level is $\left|102s_{1/2}\right\rangle$, excitation duration is $\tau=1$ $\mu$s.  Solid curves are fits of the form $P=\frac{1}{2}Ae^{-a\theta^2}(1- e^{-b \theta^2}\cos(\sqrt{N_e}\theta))$, see Methods section. The fit parameters $(A,a,b,N_e)$ are: $(4.3,1.43,1.70,456)$  for {\bf a}, $(4.44,1.43,1.77,397)$  for {\bf b}, $(3.24,1.14,0.72,243)$  for {\bf c} and $(2.56,0.79,0.86,148)$ for {\bf d}. (e) $\sqrt{N_{e}}$ as a function of number of atoms $N_a$ determined from fluorescence measurements. The data are fit with a function $C\sqrt{N_{a}}$, with the best-fit value $C=0.74$. The inset shows a collective Bloch vector tipped by the angle $\sqrt{N_e}\theta$ on the unit sphere corresponding to the many-atom states $|G\rangle$ and $|R\rangle$. The error bars represent $\pm$ one standard deviation ($\sqrt{M}$) for $M$ photoelectric counting events. }}
\label{fig:g2}
\end{figure*}

When an atom is promoted into a Rydberg level with principal quantum number $n$, the valence electron is in an orbit that is $\sim n^2$ larger than that of the ground-level atom. The atomic dipole moment is correspondingly larger, so that the dipole-dipole interaction of two atoms is increased by $\sim n^4$ \cite{gallagher}. For $n\simeq 100$ the interactions are sufficiently strong that for two atoms separated by a distance $\sim 10$ $\mu$m the associated energy shift may prevent the second atom from being excited. The excitation blockade gives rise to an oscillation between the collective ground state $|G\rangle \equiv \prod^N_{i=1}|g\rangle_i$ and the state $|R\rangle \equiv 1/\sqrt{N}\sum^N_{i=1}|g\rangle_1...|r\rangle_i...|g\rangle_N$ in which one of the $N$ atoms is in the Rydberg level $|r\rangle$, with frequency $\sqrt{N}\Omega$ \cite{lukin,allen,saffman,stanojevic,cummings}. The average number $\langle N\rangle_r$ of atoms in level $|r\rangle$ is given by:
\begin{equation}
\langle N\rangle_r=\sin^2(\sqrt{N}\Omega t).
\end{equation}
The result holds for an inhomogeneous distribution of atom-light coupling $\Omega_i$ with a modification $\sqrt{N}\Omega \rightarrow \sqrt{\sum^N_{i=1}\Omega_i^2}$ and  $| R \rangle \rightarrow (1/\sqrt{\sum_{i = 1}^{N} \Omega_i^2}) \sum_{i = 1}^{N} \Omega_i |g \rangle_1...| r \rangle_i ...| g \rangle_N$, see Methods section. For two atoms the Rydberg blockade and the accompanying $\sqrt{2}$ enhancement of the Rabi oscillation frequency have been observed by Urban {\it et al.} \cite{urban} and Gaetan {\it et al.} \cite{gaetan}. Over the past decade there have been numerous studies of the many-atom Rydberg blockade, however none of them, without exception, demonstrated either a blockade by a single atom or the many-body Rabi flopping \cite{swm,viteau,singer,vogt,tong,liebisch,heidemann,dudin}.

Here we report observations of many-body Rabi oscillations of a mesoscopic ($a\simeq 15$ $\mu$m) ensemble of rubidium atoms in the
 regime of the Rydberg excitation blockade by just one atom. In order to achieve this, the interaction strength $\Delta E_B \equiv \Delta_{ij}(a)$ between a pair of atoms at a distance equal to the ensemble size $a$ must be greater than the spectral width $\delta \omega$ of the exciting laser field. The duration of coherent atom-light interaction is limited to $\lesssim 2$ $\mu$s by the finite coherence time of the ground-Rydberg transition caused by atomic motion \cite{dudin}. Therefore, we choose a high-lying Rydberg level $|r\rangle=|102s_{1/2}\rangle$, for which $\Delta E_B\gtrsim 5$ MHz is sufficiently large.

A gas of $^{87}$Rb atoms of temperature $T \simeq 10$ $\mu$K and of peak density $\rho_0 \simeq 10^{12}$ cm$^{-3}$ is prepared in an optical lattice. The lattice is shut off, and the atoms are driven in resonance between the ground $|g\rangle=|5s_{1/2}\rangle$ and a Rydberg $|r\rangle$ level with the two-photon Rabi frequency $\Omega({\bf r}) = \Omega_1({\bf r}) \Omega_2({\bf r})/ (2\Delta_1)$ for a duration $\tau =1$ $\mu$s, with the corresponding single-atom excitation pulse area $\theta\equiv \Omega(0) \tau$, Figure 1 (a) and (b). The transverse size (Gaussian waists $w_x \approx w_y \simeq 6$ $\mu$m) of the Rydberg excitation region is determined by the overlap of the nearly counter-propagating two-photon excitation laser fields $\Omega_1$ at 795 nm and $\Omega_2$ at 475 nm. The longitudinal extent of the ensemble is determined by the sample size of waist $w_z \approx 11$ $\mu$m along $z$.

 \begin{figure}[bt]
\includegraphics[scale=1.2]{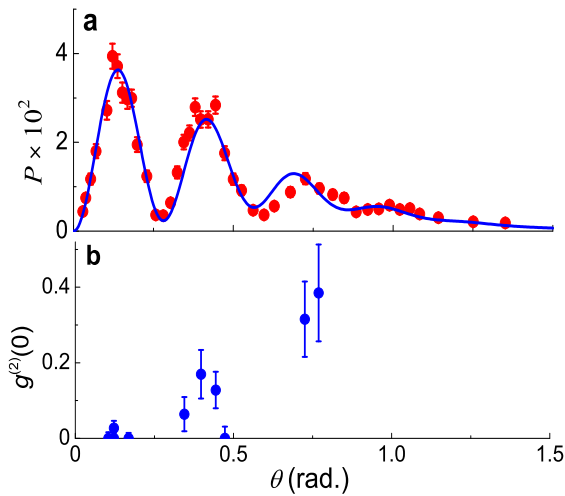}
\caption{\sffamily{ \textbf{Probability of photoelectric detection  $P$ and second-order intensity correlation function at zero delay $g^{(2)}(0)$ as a function of the single-atom Rabi angle $\theta$.}  Excitation duration is $\tau=1$ $\mu$s and upper level is $\left|102s_{1/2}\right\rangle$.  In panel {\bf a} the solid curve is a fit as in Figure 2({\bf a-d}). The fit parameters $(A,a,b,N_e)$ are $(3.80,1.48,1.86,492)$. The error bars represent $\pm$ one standard deviation ($\sqrt{M}$) for $M$ photoelectric counting events.}}
\label{fig:g2}
\end{figure}

The measurement of the population of state $|r\rangle$ is achieved by the quantum state transfer onto a retrieved light field using a 1 $\mu$s long read-out field $\Omega_3$ at 475 nm, in resonance with the $|102s_{1/2}\rangle \leftrightarrow |5p_{1/2}\rangle$ transition \cite{fleischhauer,matsukevich}.  The retrieved field is coupled into a single mode fiber followed by a beam splitter and a pair of single-photon detectors D$_1$ and D$_2$. 
Figure 2(a) shows the sum of the photoelectric detection event probabilities at the two detectors $P\equiv p_1+p_2$ as a function of the single-atom Rabi angle $\theta$, varied by changing the intensity of the $\Omega_1$ field. The data are fit with the sinusoidal oscillation of Eq. (1) modified by two Gaussians, as described in the Methods section. The choice of the fit function is motivated by a physical picture in which the visibility of the oscillation is smeared by fluctuations of the atom number and the intensities of laser fields $\Omega_1$ and $\Omega_2$, while the leading contribution to the overall decay of the retrieved signal, for the case when $\sqrt{N_e}\Omega_0<\Delta E_B$, is due to an inhomogeneous distribution of light shifts for atoms in state $|R\rangle$, $\sim N_e \Omega_0^2/\Delta E_B $ which couple the state $| R \rangle$ to other singly-excited states. The effective number of atoms $N_e$ is defined as $N_e\equiv \sum_{i = 1}^{N} \Omega_i^2/\Omega^2(0)$. For our experimental geometry $\Omega_i^2=\Omega^2(0)\exp(-2x^2/w_x^2 -2y^2/w_y^2)$, and the atom density $\rho=\rho_0\exp(-2z^2/w_z^2)$. Therefore
$N_e=(\pi/2)^{3/2}w_xw_yw_z\rho_0$.

 \begin{figure}[bt]
\includegraphics[scale=1.2]{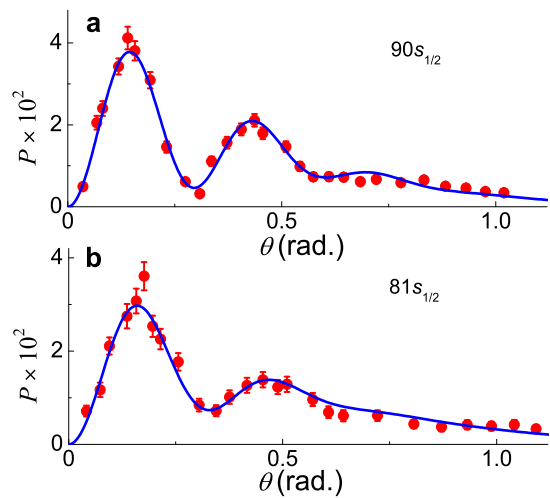}
\caption{\sffamily{ \textbf{Probability of photoelectric detection $P$ as a function of the single-atom Rabi angle $\theta$.} Excitation duration is $\tau=1$ $\mu$s. The solid curves are fits as in Figure 2({\bf a-d}), where the fit parameters $(A,a,b,N_e)$ are $(4.10,2.00,3.52,441)$ for n=90 in {\bf a} and  $(3.42,1.62,6.70,335)$ for n=81 in {\bf b}, respectively. The error bars represent $\pm$ one standard deviation ($\sqrt{M}$) for $M$ photoelectric counting events. }}
\label{fig:g2}
\end{figure}

To explore the collective character of the observed Rabi flopping, we measure $P$ as a function of $\theta$ while varying the peak density of the sample $\rho_0$, Fig. 2(b-d). Figure 2(e) shows the normalized frequency of the Rabi oscillation $\sqrt{N_e}$ extracted from the data in Figure 2(a-d) as a function of number of atoms in the ensemble $N_a$. The latter is calculated using peak density $\rho_0$ measured by the hyperfine state-selective fluorescence imaging of the atomic sample with magneto-optical trap cooling beams used without a repumping field to exclude contribution of $|5s_{1/2},F=1\rangle$ atoms. The absence of additional peaks in Fig. 1(c) supports a near-unity value for the fraction of atoms $f$ in the $m=0$ Zeeman sub-level. Ideally, we expect the effective atom number $N_e$ extracted from the Rabi oscillation period to be equal to the atom number $N_a$ determined by the fluorescence imaging of the sample, so that parameter $C$ in the fit in Fig. 2(e) would equal unity, whereas we extract $C=0.74$. In addition to the factor $\sqrt{f}$, likely cause for $C<1$ are alignment imperfections, uncertainties in the determined waists of the two-photon excitation laser beams, and uncertainties in the fluorescence measurements of $\rho_0$.

 \begin{figure}[tbh]
\includegraphics[scale=1.2]{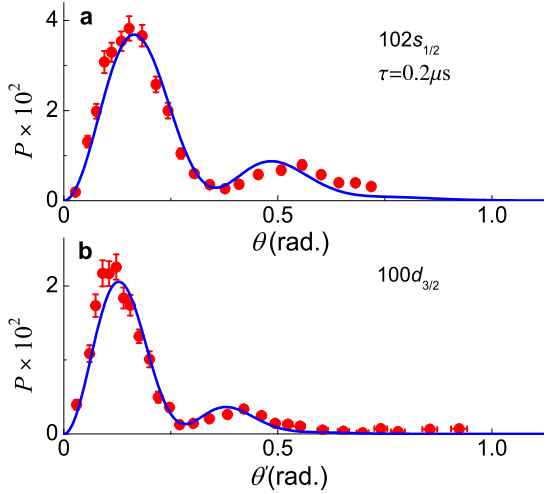}
\caption{\sffamily{ \textbf{Probability of photoelectric detection  $P$ as a function of the single-atom Rabi angle $\theta$.} Level $|102s_{1/2}\rangle$ is excited for $\tau=0.2$ $\mu$s in {\bf a}, and level $|102d_{3/2}\rangle$ is excited for $\tau=1$ $\mu$s in {\bf b}.  The solid curves are fits as in Figure 2({\bf a-d}), the fit parameters $(A,a,b,N_e)$ are $(4.56,5.27,3.86,340)$ in {\bf a}. The excitation spectrum
for $|102d_{3/2}\rangle$ shows a complex structure due to an interplay of the bias magnetic field and an ambient electric field. For the data in {\bf b}, the laser is tuned to the strongest spectral component, with the scale $\theta'$ determined by using the value of $N_e=492$ from the preceding measurements with the $|102s_{1/2}\rangle$ level, with a fit providing the value of peak single-atom Rabi frequency $\Omega_0$ and the fit parameters $(A,a,b)$ are $(2.58,10.7,3.49)$. The vertical error bars represent $\pm$ one standard deviation ($\sqrt{M}$) for $M$ photoelectric counting events. The horizontal error bars in {\bf b} reflect the uncertainty in determination of the x-axis scale $\theta'$.}}
\label{fig:g2}
\end{figure}
We further confirm that the dynamics seen in Figure 2 correspond to the oscillation of Eq. (1) by measurements of the second-order intensity correlation function at zero delay $g^{(2)}(0)$ as a function of $\theta$, shown in Figure 3. Measured values of $g^{(2)}(0)$ well below unity, together with substantial visibility of the oscillations, indicate that only one Rydberg excitation is present in the entire ensemble of several hundred atoms. The measured values of $g^{(2)}(0)$ for $\sqrt{N_e}\theta \geq 5\pi$ in Figure 3(b) suggest that the contribution of populated doubly-excited states to extracted values of $a$ is substantial. On the other hand, the sizable oscillation visibilities in Figure 2 indicate small population of the doubly-excited states for $\sqrt{N_e}\theta \approx \pi$, as the latter are expected to have a suppressed retrieval efficiency. Combining all the data points for $\sqrt{N_e}\theta \approx \pi$ in Fig. 3(b) we obtain $g^{(2)}(0)=0.006(6)$, which to our knowledge is the lowest value for this quantity for any previously reported light source. It is consistent with a lower bound of $g^{(2)}_{bg}(0)=0.012(2)$ due to background counts, of which about half are due to detector dark counts.

The importance of the condition $\Delta E_B \gg \delta \omega$ for observation of many-body Rabi oscillations is checked by reducing $\Delta E_B$ in measurements with $n=90$ and $n=81$, Fig. 4. Figure 5(a) shows data with increased $\delta \omega$ by using shorter $\tau=0.2$ $\mu$s excitation. The oscillation visibility is clearly lower both for smaller $\Delta E_B$, Figure 4, and larger $\delta \omega$, Figure 5(a). Figure 5(b) shows an oscillation with a similarly reduced visibility in measurements with the $|100d_{3/2}\rangle$ level with $\tau=1$ $\mu$s excitation, which may be attributed to the blockade breakdown due to a strong angular dependence of the atomic interaction strengths for $|nd\rangle$-levels \cite{swm}. It should be also noted that for a Gaussian distribution of atom-field couplings, describing a laser beam exciting the extended gas in our experiment, the single-atom Rabi oscillations are almost completely washed out \cite{deiglmayr}, which makes the observation of many-atom oscillations under these conditions even more remarkable.

We have demonstrated coherent many-body Rabi oscillations in an ensemble of several hundred cold rubidium atoms. The oscillations provide compelling evidence for the achievement of a collective Rydberg excitation blockade by a single excited atom. Our results pave the way for the realization of quantum simulators which employ qubits encoded in atomic ensembles, with fast two-qubit quantum gates mediated by strong Rydberg interactions \cite{lukin,weimer}.

METHODS

\scriptsize{

\textbf{Atom preparation and laser excitation.} A magneto-optical trap of $^{87}$Rb is loaded from background vapor for 70 ms. During the following 25 ms, the detuning of cooling light is increased, the repumper intensity is decreased, and the optical lattice is turned on. The lattice is composed of a single
782 nm retro-reflected linearly polarized Gaussian beam. Untrapped atoms are allowed to fall away from the experimental region during the next 15 ms period, and a $B_0=4.3$ G bias magnetic field is turned on. The trapped atoms are optically pumped
to the $|5s_{1/2}, F=2, m_F=0\rangle$ state, the optical lattice is switched off by an acousto-optical modulator (AOM), and a
3 $\mu$s long sequence of two-photon Rabi driving and retrieval is repeated for 50 $\mu$s, with a 1 $\mu$s optical pumping period included every five cycles. The overall repetition rate of the experiment is $\approx8$ Hz. For the data in Figure 2 the peak density $\rho_0$ was controlled by varying the time period between lattice loading and the two-photon excitation sequence between 15 and 90 ms. Both $\Omega_1$ (at 795 nm) and $\Omega_2$ (at 475 nm) fields are linearly polarized along the same axis. The 795 nm field is produced by an extended cavity diode laser (ECDL). Light at 475 nm is generated by frequency-doubling the output of a tapered amplifier driven by a 950 nm ECDL laser. Both lasers are frequency-locked to a thermally stabilized ultra-low expansion glass cavity and have linewidths $<100$ kHz. The transition is located by scanning the laser frequency across a resonance and measuring the photoelectric detection probability for the retrieved field. The 795 nm and 475 nm excitation fields are tuned to the two-photon resonance between the ground-level component $|5s_{1/2}, F=2,m_F=0\rangle$ and a Zeeman component of the Rydberg level $|ns_{1/2}, m_j=-1/2\rangle$.
The single-photon Rabi frequency on the blue transition is $\Omega_2 = -e \mathcal{E} \langle 5p_{1/2},F=1,m_F = 0 | r |ns_{1/2},m_j \rangle$, where $\mathcal{E}$ is electric field amplitude. The radial matrix element is reduced using the Wigner-Eckart theorem. The angular part is calculated following Ref. \cite{walker}, while the reduced matrix element is approximated by $\langle r \rangle = 0.14 \times (50/n)^{3/2}a_0$ \cite{swm}.

Since $\Omega_2$ and $\Omega_3$ fields are propagating in the same spatial mode,
the retrieved field is phase matched into the
mode of the $\Omega_1$ field and coupled into a single mode 50/50 fiber
beamsplitter followed by a pair of  single-photon detectors D$_1$ and D$_2$. A gating AOM at the fiber beamsplitter input port is employed to avoid damaging the single photon detectors by the $\Omega_1$ field.

\textbf{Photoelectric detection and data analysis.} For every experimental trial, photoelectric events on detectors D$_1$ and D$_2$ are recorded within a time interval determined by the length of the retrieved pulse. Photoelectric detection probabilities for both detectors are calculated as $p_{1,2}=N_{1,2}/N_0$, where $N_{1,2}$ are numbers of recorded events and $N_0$ is the number of received triggers. The photoelectric detection probability for double coincidences is calculated as $N_{12}/N$, where $N_{12}$ is a total number of simultaneous clicks on both detectors for a given experimental trial.
The second order intensity correlation function at zero delay is given by $g^2(0) = p_{12}/(p_1p_2)$. Transmission through the glass vacuum chamber is 0.92, the gating AOM diffraction efficiency is 0.7, the fiber coupling efficiency is 0.73, and the quantum efficiency of the single-photon counters is $0.55$. Normalizing the probability of photoelectric detection event per trial $P(\theta \approx \pi/\sqrt{N_e}) \simeq 0.04$ by the product of these efficiencies we arrive at the single-photon generation efficiency into the Gaussian mode of our single-mode fiber $\epsilon=0.15$.

\textbf{Decoherence model.} We employ the following Hamiltonian to describe our system: $\hat{H} = \sum_{\mu} \hbar (\omega_{g} \hat{\sigma}_{\mu}^{gg} + \omega_{r} \hat{\sigma}_{\mu}^{rr}) + \frac{1}{2}\sum_{\mu} \hbar (\Omega_{\mu} e^{-i \omega_L t} \hat{\sigma}_{\mu}^{rg} + h.c.) + \sum_{\mu > \nu} \hbar \Delta_{\mu\nu} \hat{\sigma}_{\mu}^{rr} \otimes \hat{\sigma}_{\nu}^{rr}$. The atomic operators for the atom $\mu$ are defined as $\hat{\sigma}_{\mu}^{ab} = | a \rangle_{\mu}\langle b |$, where $a,b \in [g,r]$ with $| g \rangle_{\mu}$ being the atomic ground state and $| r \rangle_{\mu}$ being the addressed Rydberg level. The two-photon excitation is modeled using the effective Rabi frequency $\Omega = \Omega_1 \Omega_2 / (2\Delta)$. The interaction between Rydberg levels is described with a single-channel model.For $\Delta_{\mu\nu} \gg \Omega_{\mu},\Omega_{\nu} \;\forall (\mu,\nu)$, the excitation blockade is operational. Adiabatic elimination of double and higher-order excitations from the equations of motion results in an effective Hamiltonian for the singly-excited part of the spectrum: $\hat{H}_{eff} = \sum_{j} \hbar \Delta_{j} | j \rangle\langle j| + \sum_{i > j} \hbar C_{ij} (| i \rangle\langle j| + | j \rangle\langle i|) +
\frac{1}{2}\sum_{j} \hbar \Omega_{j} (| j \rangle\langle G| + | G \rangle\langle j|)$. Here $\Delta_{j} = - \sum_{i \neq j} \Omega^2_{i}/(4 \Delta_{ij})$, $C_{ij} = - \Omega_{i} \Omega_{j} / (4 \Delta_{ij})$, where $| j \rangle$ is the many-body state with the $j$-th atom in the Rydberg level. The first two terms of the effective Hamiltonian are due to the light shifts induced by the (detuned) doubly-excited states onto the single excitations.

When the interaction-induced inhomogeneous light shifts are omitted, the Hamiltonian results in an ideal Rabi oscillation between the ground state $|G\rangle$ and the single spin wave $|R\rangle = (1/\sqrt{\sum_{j} \Omega_{j}^2}) \sum_{j} \Omega_{j} | j \rangle$. If at time $t=0$ the system is in state $| G \rangle$, the state at future times is given by $| \psi(t) \rangle = \cos(\Omega t/2) |G\rangle - i \sin(\Omega t/2) |R\rangle $.
When the light shift terms are included, the state $|R\rangle$ is coupled to a broad distribution of singly-excited states and therefore leaks into this quasi-continuum, leading to $P\sim |\langle R|\psi(t) \rangle|^2$ decaying with a rate $\sim N_e\Omega_0^2/\Delta E_B$. 
The doubly-excited states are expected to be populated at a rate $\sim N_e\Omega_0^2$. Trial-to-trial fluctuations $\Delta\Omega$ and $\Delta N_e$ in $N_e$ and $\Omega_0$, respectively,  lead to a decay of the oscillation visibility. The probability of photoelectric detection per trial $P$ as a function of $\theta$ in Figures 2-5 is therefore fit by a function $\frac{1}{2}Ae^{-a\theta^2}(1- e^{-b \theta^2}\cos(\sqrt{N_e}\theta))$, where dimensionless fit parameters $a\sim N_e$ and $b\sim (\Delta\Omega/\Omega_0)^2+(\Delta N_e/2N_e)^2$ describe the roles of the light shifts and population of doubly-excited states, and atom number and intensity fluctuations, respectively, while an amplitude $A$ represents the overall measured retrieval and detection efficiency. }

 {\small \bf Acknowledgments.} {\small
We thank A. Radnaev and B. Kennedy for discussions. This work was supported by the Atomic Physics Program and the Quantum Memories MURI of the Air Force Office of Scientific Research and the National Science Foundation.}

{\small \bf Author information.} {\small Correspondence and requests for materials should be addressed to A.K.}

\end{document}